# LLM and Agent-Driven Data Analysis: A Systematic Approach for Enterprise Applications and System-level Deployment

Xi Wang[*1,2]    Xianyao Ling[2]    Kun Li[2]    Gang Yin[2]    Liang Zhang[1,3]
Jiang Wu[3]    Annie Wang[4]    Weizhe Wang[5]

[1]Tsinghua University    [2] Cross-strait Tsinghua Research Institute
[3]OceanBlue Construction Co. Beijing, Ltd.    [4]Kalavai Corp
[5]Beijing Eastern Golden Info Technology Co., Ltd.

xi-wang19@mails.tsinghua.edu.cn    xianyao.ling@ctri.org.cn    likun@ctri-ai.cn
yingang@ctri-ai.cn    liang-zh19@mails.tsinghua.edu.cn
jiang.wu@rcytgs.com    annie@kalavai.net    wangwz20@mails.tsinghua.edu.cn

**Abstract:** The rapid progress in Generative AI and Agent technologies is profoundly transforming enterprise data management and analytics. Traditional database applications and system deployment are fundamentally impacted by AI-driven tools, such as Retrieval-Augmented Generation (RAG) and vector database technologies, which provide new pathways for semantic querying over enterprise knowledge bases. In the meantime, data security and compliance are top priorities for organizations adopting AI technologies. For enterprise data analysis, SQL generations powered by large language models (LLMs) and AI agents, has emerged as a key bridge connecting natural language with structured data, effectively lowering the barrier to enterprise data access and improving analytical efficiency. This paper focuses on enterprise data analysis applications and system deployment, covering a range of innovative frameworks, enabling complex query understanding, multi-agent collaboration, security verification, and computational efficiency. Through representative use cases, key challenges related to distributed deployment, data security, and inherent difficulties in SQL generation tasks are discussed.

## 1. Introduction

The rapid development of artificial intelligence particularly in large language models (LLMs), have significantly influenced enterprise data infrastructure and applications [1]. On the one hand, software development tools such as Cursor are fundamentally changing software coding practices via natural language interaction [2]. On the other hand, enterprise data analysis is evolving from structured data storage, such as data warehouses into dynamic knowledge sources that support natural language querying. In this context, text-to-SQL which can automatically translating natural language queries into structured query language (SQL), has emerged as a critical enabling technology [3]. In addition, ontology-based knowledge generation becomes increasingly focused research aera [4].

Against the backdrop of accelerating enterprise digital transformation, business users have a surging demand for real-time data analysis. However, traditional SQL coding is error-prone and relying on specialized skills, which becomes a major bottleneck in data utilization. LLMs which have powerful semantic understanding and generation capabilities, offer the potential to bridge the semantic gap between natural language and database schemas [5]. Furthermore, combined with agent architectures, systems can achieve multi-turn interactions, self-correction of errors, context awareness, and security policy enforcement, thereby building truly usable enterprise-level data dialogue systems [6]. These frameworks continuously enhance performances benchmarks like Spider [7] and BIRD [8].

---

[*] Corresponding author.

Simultaneously, enterprise data security and compliance are becoming increasingly demanding. Retrieval-Augmented Generation (RAG) technology, by embedding database schemas into vector spaces, effectively supports semantic retrieval [9]. Nevertheless, SQL generation for enterprise databases continues to face several key challenges, including the ambiguity of natural language, the complexity of database schemas, the unpredictability of SQL execution outcomes, and the limited coverage of real-world enterprise scenarios in existing evaluation benchmarks.

This paper provides a systematic approach for enterprise applications and system-level deployment of SQL generations. Representative SQL generation frameworks from recent years are presented, emphasizing technology approaches and performance characteristics. The discussion subsequently focuses on application examples in vertical domains and other representative enterprise scenarios. Finally, the limitations of current approaches are discussed, and key capabilities required for enterprise intelligent data agents are presented, relating to security validation, automatic repair, resource efficiency, and multi-domain generalization.

As shown in Figure 1, the overall framework of enterprise-level knowledge generation system in this paper consists of four layers: (1) Infrastructure Layer: Provides the foundational computing resources (GPU servers, storage device) and distributed system deployment platforms (MoPaaS, Kalavai) necessary for running AI models and GPU systems; (2) Data Layer: Manages diverse enterprise data sources, including structured data (databases, data warehouses) and unstructured data (text, network data), typically stored in distributed systems; (3) Knowledge Generation and Agent Layer: This is the core layer, which uses LLMs and multi-agent collaboration for key tasks, such as understanding the database structure (schema linking), generating SQL queries from natural language, validating results using knowledge graphs, and documents and internet data; (4) Application Layer: Delivers the final user-facing capabilities, emphasizing a natural language interface like AI Coding, comprehensive analysis for decision support, while ensuring enterprise-grade data security and compliance. These layers will be introduced in detail in the following sections. Specifically, Section 2 and 3 focus on state-of-art AI/SQL development methodologies and best practices; Section 4 and 5 focus on applications and system deployments.

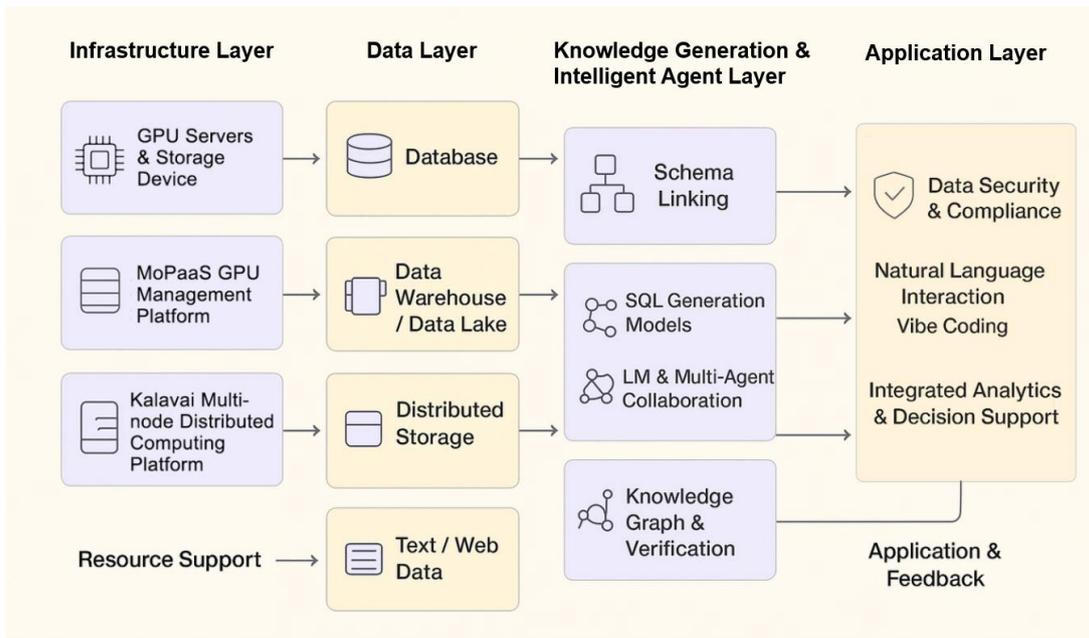

Figure 1. Overall framework of enterprise-level SQL generation system.

## 2. Enterprise Data Preparations for LLM-Based Knowledge Generations

The effectiveness of LLMs in enterprise knowledge generation systems depends on the structure, semantics, and preparation of organizational data. As majority of the corporate data are stored in SQL databases, this section introduces a unified framework that enables LLMs to generate SQL accurately by integrating three capabilities: (1) cross-source data governance and semantic alignment, (2) schema enrichment through natural-language descriptions, and (3) context optimization via Retrieval-Augmented Generation (RAG). These processes collectively transform heterogeneous data into an AI-interpretable semantic layer, providing a foundation for scalable and intelligent SQL generation in enterprise environments.

### 2.1 Enterprise Data Sources and Infrastructure

In enterprise-level SQL generation systems, the data sources serve as the foundation for intelligent querying and semantic understanding, providing high-quality, accessible, and structured knowledge to SQL generation models. Enterprises typically manage multi-source and heterogeneous data assets, including relational databases (such as operational and historical logs), data warehouses (for cross-departmental integration and analytical aggregation), distributed data lakes (for structured, semi-structured and unstructured data), as well as external web data sources such as open knowledge bases, web pages, APIs, and industry data feeds. These external resources enhance the temporal validity, domain coverage, and contextual reasoning capacity of the SQL generation system.

Given the heterogeneity in structure, semantics, and access policies across data sources, enterprises use a unified data preprocessing and semantic integration framework to transform raw information into AI-interpretable knowledge. This framework consists of four main stages: data cleaning, structural extraction, metadata normalization, and schema alignment. The system first removes redundant entries, repairs missing values, and standardizes formats. Then, convert unstructured web and document data into indexable tables or knowledge fragments. Finally, knowledge graph enhancement and schema linking establish a unified ontology semantic layer that allows AI models to perform cross-source reasoning and retrieval.

As the foundation of the data ecosystem, the infrastructure provides the necessary computational and storage capabilities. GPU clusters, cloud-native storage systems, and distributed computing platforms empower the model to perform large-scale training and inference efficiently. Meanwhile, resource scheduling and fault-tolerant mechanisms enhance scalability and service continuity. The data sources and infrastructure together form a synergistic foundation that enables enterprise SQL generation systems to achieve intelligent data interaction.

### 2.2 Database Description Generation: Building Semantically Rich Context

Schema linking aims to establish accurate semantic mappings between natural language queries and the underlying database schema [10]. The primary purpose is to identify relevant tables, columns, primary/foreign key relationships, and business entities from user queries, and to correctly incorporate them into the generated SQL statements. Recently, the integration of LLMs with RAG has become increasingly prevalent [11]. As shown in Figure 2, the schema linking pipeline leverages LLMs to transform database metadata into semantically enriched schema representations, thereby providing rich context for RAG.

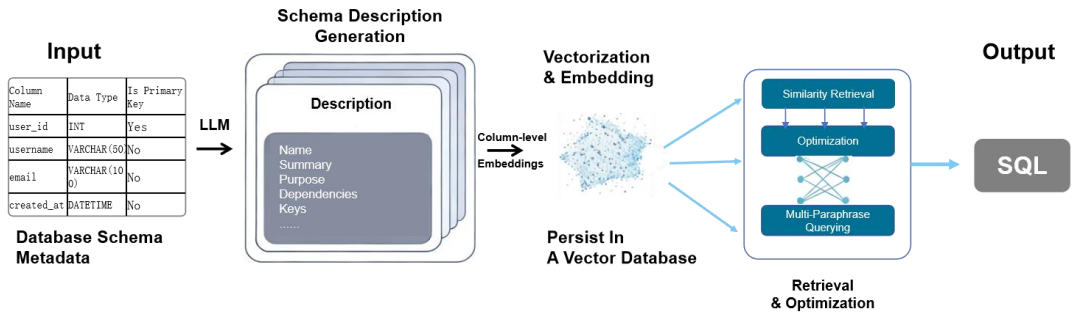

Figure 2. The evolution of schema linking from metadata parsing to RAG-enhanced semantic retrieval with domain knowledge integration (inspired by [12][13]).

To generate SQL, LLM must first connect to the enterprise database and understand its structure, which is a challenge considering the complex schema and numerous tables involved. Providing natural language descriptions of the schema can improve the model's semantic understanding. Because metadata like table and column names offer little semantic clarity, researchers suggest using automatically generated schema descriptions as "contextual prompts" to guide LLM reasoning.

The M-Schema is a representative work in this direction [12]. The structured representation provides a practical framework for modeling database semantics. The Datrics SQL system generate multiple candidate sets by calling the LLM iteratively for the same table. Then, it filters out repeated entities through frequency-based aggregation, which encodes structural information to enhance semantic depth [13]. These rich descriptions serve not only as static context, but also as later retrieval indexes, significantly improving the recall accuracy of schema elements.

## 2.3 RAG-Augmented SQL Generation

In large-scale enterprise database environments, data schemas are often very complex, such as a large number of tables, non-standardized column names, missing or redundant metadata, and implicit business semantics in the context [14]. In this case, directly inputting the complete schema into LLM will significantly increases its contextual burden.

Therefore, both the complexity of the input context and its relevance to the user's query directly determine the effectiveness of SQL generation. On the one hand, RAG compresses the input context through vector retrieval mechanisms, enabling efficient filtering of massive schema elements. On the other hand, it incorporates additional domain knowledge and SQL template, providing auxiliary information to support the LLM's reasoning and generation processes. This "on-demand supply" strategy for schema linking effectively alleviates the problem of cognitive overload in LLMs.

### 2.3.1 Vectorization and Retrieval of Schema Elements

A key factor in implementing RAG for schema retrieval lies in constructing semantic vector representations of database schemas [15]. Fine-grained and column-level embedding methods enhance retrieval accuracy by encoding each column as a dense vector. The input features for these embeddings include not only the column name but also its associated table name, business descriptions, and representative value examples [16][17]. This multi-source fusion strategy enriches the representation of business semantics.

All column embeddings are computed during an offline preprocessing phase and stored in a vector database. During inference, the same encoder is used to map user queries to query vectors and retrieve the top-k most relevant columns based on cosine similarity [18][19]. This approach outperforms coarse-grained strategies that rely solely on table names or keyword matching on benchmarks such as Spider and BIRD, particularly in scenarios where column names are ambiguous or commonly abbreviated [20].

Nevertheless, relying solely on similarity-based ranking still introduces redundant or suboptimal schema elements, such as retrieving multiple semantically similar columns simultaneously, or including irrelevant tables due to high-frequency keywords. To alleviate this issue, for an example, KaSLA selects a subset of tables and columns that maximizes the overall semantic coverage of the user query under specific constraints, such as maximum context length or a limited number of columns. This effectively formulates schema linking as a combinatorial optimization problem [21].

### 2.3.2 Multi-Perspective Querying and SQL Template Library Retrieval

Recently, SQL template knowledge bases have proven beneficial for SQL generation. This approach standardizes historical "question-SQL" pairs into structured JSON objects, which contain fields such as the original question, extracted entities, data sources and aggregation logic, and stores them in a vector database. During the generation phase, the system first retrieves semantically similar historical examples as code-level references. These templates not only supply syntactic structures but also encourage the model to reuse domain-specific query patterns, thereby improving the success rate of complex query generation. As an example, Datrics SQL persistently stores each successful SQL generation result, including the original natural language question, the generated SQL, the data tables involved, execution context, and verification status. This allows the system to continuously learn from historical experience rather than reasoning from scratch for each new request.

The method also introduces a multi-expression query strategy, which generates multiple semantically equivalent natural language variants for the same user request and performs vector retrieval for each variant separately. Since different expressions may activate different related tables, their intersection usually produces results with higher confidence [13].

### 2.3.3 Integration of Professional Domain Knowledge

In enterprise-level SQL generation systems, relying exclusively on lexical or sentence-level embedding similarity may lead to SQL queries with semantic inaccuracies or logical inconsistencies.

To alleviate this issue, researchers have proposed integrating structured database knowledge, such as table schemas and cell values, into LLMs through end-to-end text generation. This approach utilizes three specialized training tasks: column name semantics, table name semantics, and schema structure modeling [22].

Furthermore, business rules are maintained in a declarative form, separate from the model's parameters. This allows domain experts to directly refine the system without requiring model retraining by machine learning engineers. By explicitly integrating domain knowledge into the retrieval and generation pipeline, this mechanism helps build traceable and configurable intelligent query systems.

## 3. SQL Generation Models

In the SQL generation task, the architecture and training paradigm of the generation model directly determine the system's accuracy. With the rapid improvement of open-source models like Llama, Qwen, Phi, etc., the research focus is shifting from Prompt Engineering towards Parameter-Efficient Fine-Tuning and feedback-driven optimization. As shown in Figure 3, two primary training paradigms have emerged for SQL generation models: supervised fine-tuning and reinforcement learning.

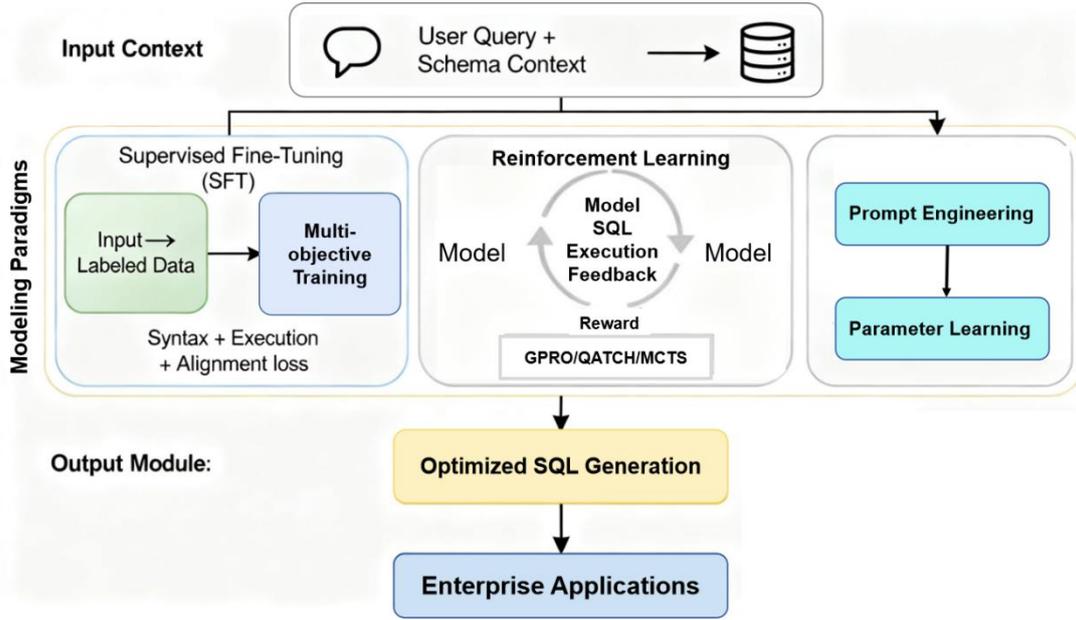

Figure 3. Overview of SQL generation models illustrating supervised fine-tuning, reinforcement learning, and the shift from prompt-based to parameter-based paradigms.

## 3.1 Supervised Fine-Tuning

Supervised Fine-Tuning (SFT) is an effective means to enhance the SQL generation capability of open-source model [23][24]. Its core idea is to train the model to learn the conditional mapping from the input question Q and database schema S to the target SQL statement through large-scale "natural language question - SQL query" pairs [25][26]. Formally, given a pre-trained language model with parameters θ, the objective function for SFT is:

$$L_{SFT}(\theta) = -\log P(SQL|Q,S;\theta)$$

Effective schema information S must extend beyond table and column names to include data types, key constraints, natural language descriptions, value examples, and table relationships. This contextual information can significantly enhance the model's understanding of complex JOINs, nested subqueries, and aggregation logic.

To improve fine-tuning effectiveness, multiple works have proposed refined training strategies. Xiyan-SQL designed a multi-objective loss function, encouraging the model to simultaneously optimize grammatical correctness, execution accuracy, and schema alignment [12]. ROUTE adopts multi-task supervised fine-tuning, jointly training SQL generation, schema linking prediction, and error detection subtasks to enhance the model's generalization ability [27]. SENSE utilizes synthetic data augmentation, generating large volumes of high-quality training samples through template perturbation and entity replacement, alleviating the scarcity of real annotated data [28]. These strategies collectively push the performance of open-source models on benchmarks like Spider and BIRD to approach or even surpass early zero-shot methods based on GPT-4.

## 3.2 Reinforcement Learning and Preference Optimization

Although supervised fine-tuning can produce syntactically correct SQL queries, models remain prone to semantic errors when handling ambiguous intents, nested logic, or complex database schemas. To improve the accuracy and reliability of generated outputs, reinforcement learning (RL) has been increasingly adopted as a complementary approach to conventional supervised learning.

Early work like Seq2SQL [29] was among the earliest to use SQL execution results as a reward signal. It decomposed SQL generation into subtasks and applied policy gradient methods to unordered "where" clauses. The reward function was defined as "+1" for correct execution, "–1" for incorrect execution, and "–2" for invalid SQL. Despite its simplicity, this approach aligned the optimization objective with final execution correctness and improved end-to-end performance on the WikiSQL benchmark.

Subsequent studies extended this direction. CogniSQL-R1-Zero [30] and Arctic-Text2SQL-R1 [31] both utilize Group Relative Policy Optimization (GRPO). This method samples N SQL candidates for each natural language question and computes relative advantages within the group based on binary execution outcomes (correct/incorrect). By leveraging group-wise comparisons, GRPO enhances training stability without complex reward shaping. On the BIRD benchmark, these frameworks achieve execution accuracies of 59.97% and 71.83%.

To address the sparsity of binary execution rewards, researchers have explored finer-grained reward designs. For instance, Think2SQL [32] incorporates dense reward signals from the QATCH evaluation framework, including cell precision, cell recall, and tuple cardinality, providing incremental feedback for partially correct SQL queries and enriching the learning signal during RL training.

Modeling SQL generation as a sequential decision process also enables integration with search algorithms. SQL-o1 [33] adopts Monte Carlo Tree Search (MCTS) as its core reasoning engine. At each step, MCTS performs exploration, expansion, simulation, and backpropagation over possible SQL tokens. It further introduces a self-reward mechanism, where the model dynamically generates reward signals based on the likelihood and quality of its own outputs, guiding the search toward more plausible and efficient solutions.

Overall, RL helps narrow the gap between model outputs and task objectives by directly linking optimization to SQL execution correctness or semantic fidelity [34][35]. The evolution from GRPO's group-based optimization and QATCH's dense rewards to MCTS-driven structured search illustrates a practical pathway for deployment in real-world database systems [36][37][38].

### 3.3 Multi-Agent Frameworks

Early SQL generation systems, such as DAIL-SQL [39], relied heavily on prompt engineering. PET-SQL uses "reference-enhanced" prompting to inject table structures and sample values, first enabling the LLM to generate initial SQL and subsequently perform more accurate schema linking. A refined prompt is then used to produce the final SQL version [40]. When handling enterprise-scale complex queries such as cross-table joins, ambiguous intent interpretation, or multi-hop reasoning, single-model architectures often struggle to simultaneously achieve high accuracy. To address this limitation, recent studies have proposed multi-agent SQL generation systems composed of specialized agents that collaborate to fulfill distinct subtasks [41][42]. These frameworks typically follow a "Think-Act-Observe" closed-loop reasoning paradigm, improving SQL generation quality in complex scenarios through role division, iterative correction, and task decomposition [43][44][45].

A typical multi-agent SQL generation system includes three roles, as shown in Figure 4.

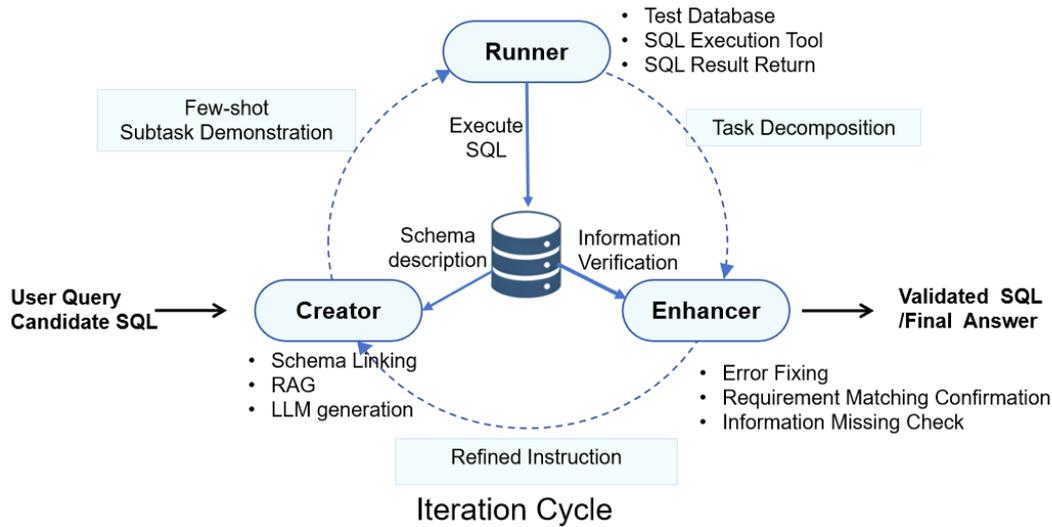

Figure 4. Feedback-driven multi-agent architecture for SQL generation. The SQL Creator, Runner and Enhancer form a clockwise iterative loop centered on the database, supported by prompting, demonstration, and task-decomposition strategies to improve query accuracy and robustness.

This process forms a feedback-driven closed loop: SQL Creator generates → Runner verifies → Enhancer enhances → Creator regenerates, until an executable and semantically correct SQL is obtained. It is noteworthy that not all enhancement steps yield performance gains. Researchers systematically studied six lightweight test-time scaling strategies and found that "Divide-and-Conquer" prompting and few-shot subtask demonstrations consistently improved the performance of both general-purpose and reasoning-focused LLMs, whereas blindly adding intermediate reasoning steps like additional planning agents or Schema filters might instead introduce noise or latency, with mixed effectiveness [46]. This indicates that the design of agent architectures requires a delicate balance between accuracy, efficiency, and complexity.

DB-GPT[47] extends the basic SQL generation agent by introducing role-based collaboration. The system defines multiple specialized roles such as data analyst, software engineer, and database architect, and associates each with a set of standardized operating procedures (SOPs) to support end-to-end database interaction workflows. By assigning different responsibilities to individual agents and leveraging their domain-specific expertise to handle complex tasks.

## 4. Knowledge Graph Enhanced SQL Generation Methods

Although LLM has made significant progress in natural language understanding and code generation, there are still limitations in relying on database schemas and natural language prompts in enterprise-level SQL generation tasks. The main challenges include:

(1) Semantic Ambiguity: The same term may have different meanings in different business contexts.

(2) Missing Implicit Constraints: User queries often lack descriptions of business logic constraints.

(3) Cross-Table Entity Linkage: Complex enterprise databases contain multi-table, multi-domain, multi-level relationships that are difficult to capture through a single schema.

To address these issues, Knowledge Graphs (KGs), as structured external knowledge sources, are introduced to enhance LLM capabilities in semantic modeling, query generation, and reasoning. As shown in Figure 5, the KG-enhanced SQL generation framework includes three stages:

knowledge extraction, graph construction, and graph-based query enhancement.

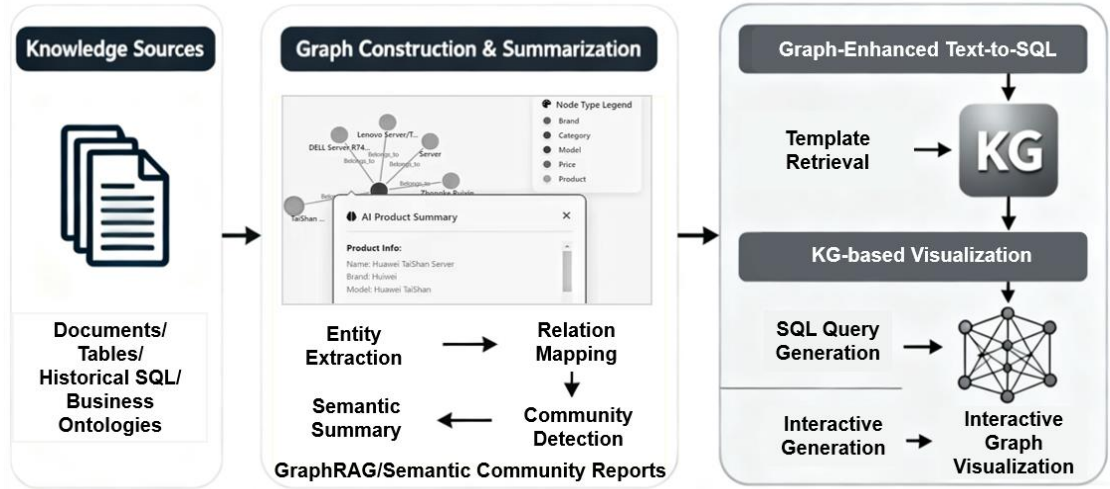

Figure 5. Integration of knowledge graphs into SQL generation to enhance reasoning, template retrieval, and result visualization (inspired by [48][49]).

### 4.1 Knowledge Graph Extraction Based on LLMs

Graph-based Retrieval-Augmented Generation (GraphRAG) is a technical framework that combines KGs with the RAG mechanism, aiming to enhance the accuracy and explainability of LLMs in question answering, reasoning, and generation tasks [49]. It is based on traditional text chunking and vector retrieval, introducing structured graph knowledge to capture complex relationships and multi-hop semantics.

The specific processes include:

Knowledge Extraction Stage: Split the source text into several semantic chunks, using the LLM to automatically extract Entities, Relations, and Descriptions. If the same entity appears in multiple text chunks, description merging, and semantic aggregation will be performed.

KG Building Stage: Build a preliminary KG based on the extracted entities and relations, and form semantic communities through community detection algorithms (like Leiden clustering).

Semantic Summarization Stage: For each community, the model generates a "Community Report", which summarizes the entity relationships and key semantics within it.

Q&A Reasoning Stage: For a user query, the system first locates relevant communities, generates local answers (Map-level Answer), and then synthesizes them into a final result through an aggregation model.

This method has significant value in enterprise data scenarios. Enterprise data is often scattered across multiple business systems such as finance, sales, supply chain, etc. Although database tables may lack explicit foreign key associations, rich implicit relationships exist at the semantic level. Through KG modeling, it can assist the model in identifying semantic associations and business logic among data during the SQL generation process. For example, when a user queries the "profit distribution of the highest sales product category across different regions", the KG can help the model automatically associate the "sales record table", "product category table" and "regional profit table", enabling automatic generation and optimization of cross-table SQL. Therefore, introducing KGs as a "structured semantic enhancement layer" in SQL generation can effectively improve the model's understanding and generation capability for complex query semantics.

## 4.2 SQL Template Library Retrieval Based on Knowledge Graph

In practical systems, enterprises often have a large amount of historical "natural language question - SQL query" paired data. By standardizing and modeling these historical data as knowledge, the accuracy and reusability of SQL generation can be significantly improved.

First, convert historical data into a structured JSON format, include Query Question, involved Entities and Data Source, Aggregation and Filtering Logic, Output Schema, etc. Then, store them in a Vector Database and build an SQL Template KG. The nodes of this graph can represent query topics, aggregation logic, business metrics, or specific SQL patterns, while the edges describe their semantic similarity or logical dependencies.

During the generation phase, when a user inputs a new natural language query, the system first computes its semantic embedding vector and retrieves the most similar query template via the graph. These templates not only provide syntactic structure information but also guide the model to reuse domain-specific query patterns such as financial ratio calculation, user segmentation, inventory analysis, etc., effectively reducing generation error rates. By constructing an SQL Template Graph, the system can also macroscopically display the thematic structure of enterprise data analysis. Cross nodes between query templates of different business units reflect potential data sharing needs. This knowledge can be further fed back to data modelers, guiding database structure optimization and query pattern reuse.

## 4.3 Visualization of SQL Query Results Based on Knowledge Graph

Beyond assisting generation and retrieval, KGs can also be used for the visualization and explanatory enhancement of SQL query results. Traditional SQL query results are usually presented in two-dimensional table forms, making it difficult to reveal hierarchical relationships and business logic between metrics. After introducing KGs, visualization enhancement can be achieved in the following ways:

(1) **Relation-Oriented Visualization:** Map SQL query results to graph nodes, generating a semantic network of "Metric - Entity – Relation".

(2) **Dynamic Query Path Tracing:** Record the query generation and execution path through the graph structure, helping users understand how the model reasoned to obtain the result, thereby enhancing the explainability and trustworthiness of the results.

(3) **Interactive Exploration:** Users can click on nodes or relations in the graph interface to trigger new SQL queries, achieving "deep semantic analysis".

Taking Figure 6 as an example, which shows a product KG in our previous work [48]. It includes the following interactive exploration functions: Expand Node, Hide Node, AI Product Analysis. As shown in Figure 6, after selecting the "Huawei TaiShan Server" product node, the user can right-click the menu and click "Expand Node". Then, the system dynamically loads and displays all primary neighbor nodes directly connected to this product node, namely its brand "Huawei", model "Huawei TaiShan" and price "23500 yuan". This function provides users with clear context awareness and in-depth analysis of query nodes.

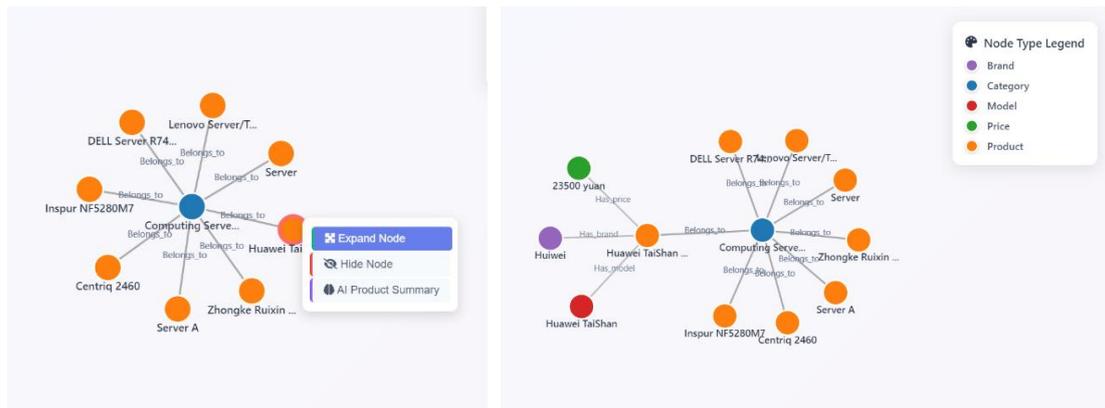

Figure 6. Example of visualization and interactive exploration based on knowledge graph (cited and modified from [48]).

In enterprise decision support systems, KG-based visualization improves the transparency of data queries, and supports an end-to-end pipeline spanning natural language querying, SQL generation, and result interpretation, thereby enhancing the usability of SQL generation systems.

The integration of KGs into SQL generation reflects a shift from "language-to-code mapping" toward ontology based "semantic-structural integration". This approach enhances the model's semantic understanding and reasoning capabilities, which helps to improve explainability, scalability, and user interaction in complex enterprise environments.

## 5. System Deployment and Applications

Although SQL generation has shown consistent improvement on academic benchmarks, its large-scale deployment in enterprise systems continues to face multiple challenges, including security assurance, output accuracy, system integrability, user experience, and the delineation of functional boundaries. As shown in Figure 7, a typical enterprise system deployment architecture adopts a three-layer design: User Interaction Layer, Intelligent Processing Layer, and Security Execution Layer. In recent years, advances in multi-agent architectures, security governance mechanisms, and emerging human-AI collaboration paradigms, like AI Coding, have facilitated the transition of SQL generation systems from research projects toward enterprise system production use. This section examines deployment methodologies for enterprise environments.

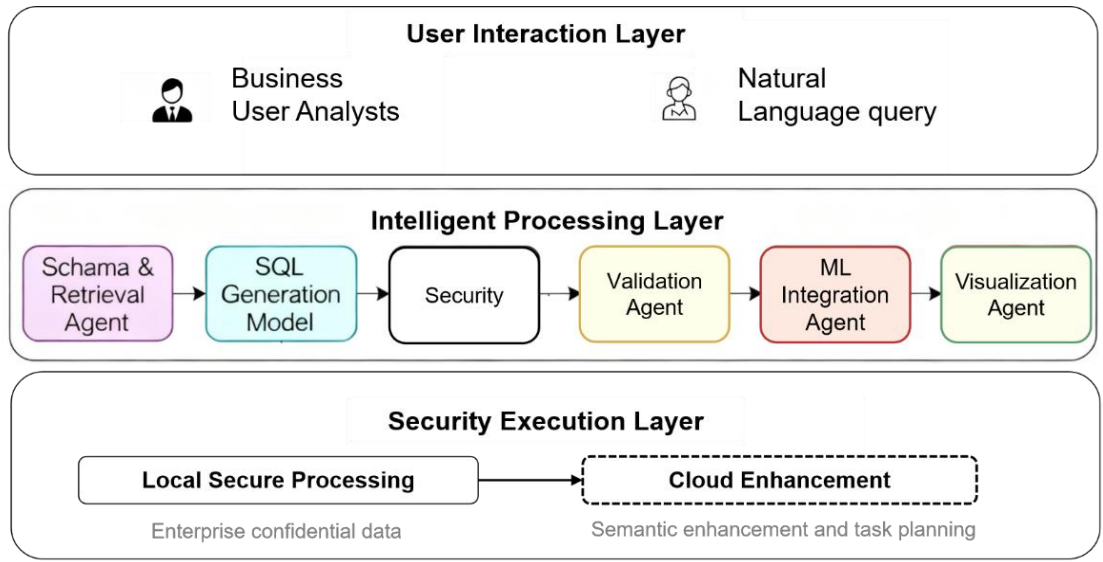

Figure 7. System deployment architecture illustrating local-cloud dual-layer security, multi-model collaboration, and integration with advanced analytical workflows.

**5.1 Data Security Based on a Dual-Layer Model Architecture**

In enterprise-level data analysis systems, SQL data processing capabilities must ensure the following: high speed, analytical accuracy, and data security. Models with larger parameter scales tend to be slower in processing but offer higher analytical accuracy. However, enterprise hardware often cannot support such large-scale models, making cloud-based LLM services an ideal solution for this need. On the other hand, models with smaller parameter scales are faster and more flexible but are typically limited to processing data with relatively homogeneous characteristics, making them more suitable for on-premises deployment.

In practice, enterprises usually choose model deployment methods based on their specific needs. However, due to data security constraints, deployment approaches often lack flexibility. To address this issue, we propose a dual-layer model architecture designed to ensure enterprise data security. This method involves data classification and masking at the data layer, along with dual-model inference based on data characteristics, to accomplish SQL generation tasks.

MoPaaS platform equips single-node machines with the ability to deploy and train LLMs, characterized by dynamic model switching under constrained VRAM. As shown in Figure 8, this functionality helps fine-tuning domain-knowledge models and data analysis models, enhancing local data processing capabilities.

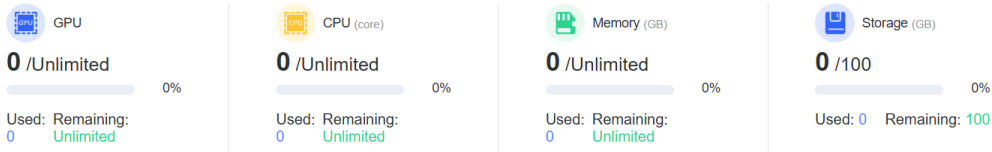

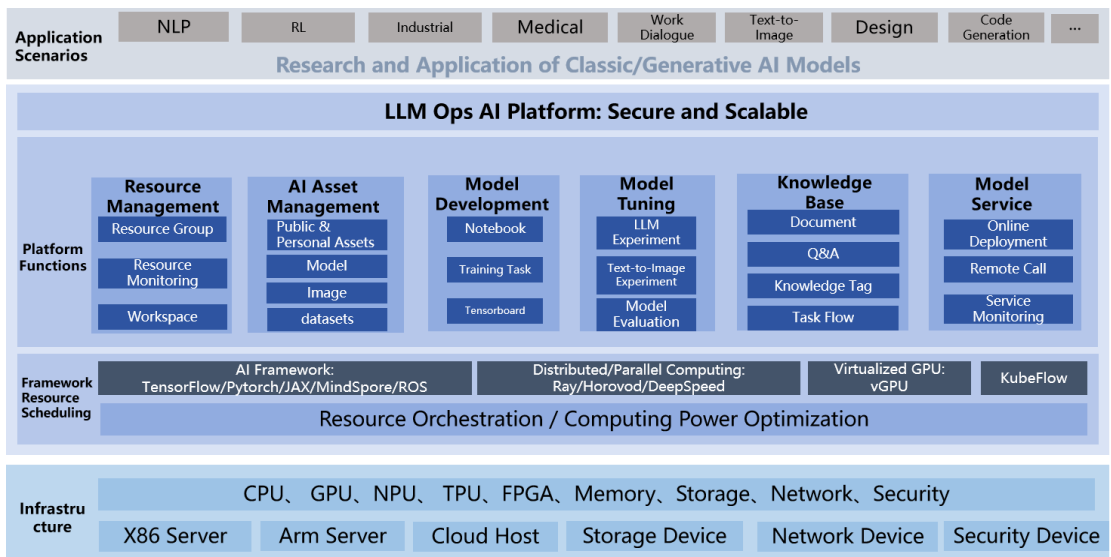

Figure 8. Example of a local system deployment platform (e.g., MoPaaS), enabling rapid on-device deployment of LLMs to mitigate potential data security risks associated with cloud-based inference and support local monitoring of the inference process.

As shown in Figure 9, Kalavai distributed computing platform constructs a distributed LLM resource pool by interconnecting multiple computing devices, including cloud instances, local servers, and personal computers, and centrally orchestrates inference workloads across them. This architecture supports multiple model engines, with processing capacity scaling according to the number of worker nodes, providing both cloud-based and on-premises LLM processing capabilities. It can also dynamically scale multiple nodes in real time based on task complexity.

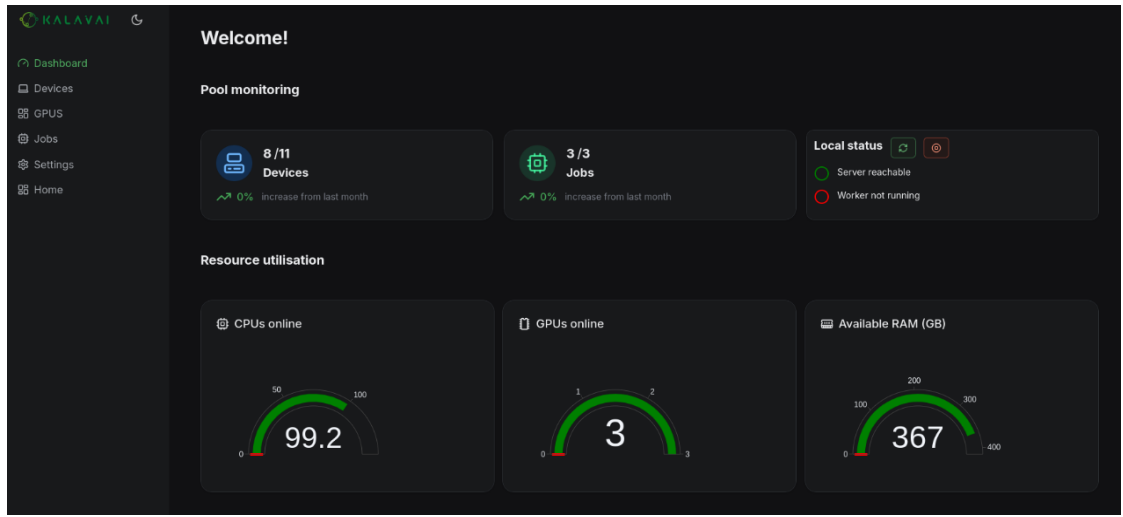

Figure 9. Dashboard interface of Kalavai multi-node computing platform.

Both aforementioned architectures provide infrastructure-level solutions for local models and multi-node deployments. Currently, models like ChatGPT, Gemini, ChatGLM, and Qwen offer cloud-based APIs. These models are supported by robust hardware infrastructure, enabling the deployment of larger parameter-scale models with faster processing speeds and enhanced instruction following capabilities. Consequently, the steps of processing textual data in the cloud often result in better performance. However, since models often interact directly with databases, inadequate security measures during such interactions may lead to accidental exposure of sensitive information or unauthorized data access. Unlike general natural language generation tasks, SQL generation involves inputs, contextual information, and outputs that may contain real business data, internal architectural elements, or implicit business constraints. Therefore, it is essential to establish clear security boundaries between model inference and underlying data access.

During the processes of generating database schemas, documents, or knowledge graphs mentioned in Sections 2.2 and 4.1, sensitive information is reviewed and semantically masked prior to LLM data processing through privately deployed models within the local system. These models do not communicate with external cloud models, transmitting only abstracted schema skeletons to cloud-based models for subsequent semantic enhancement or task planning, thereby ensuring that raw data never leaves the enterprise boundary. In the database and knowledge base retrieval phase described in Section 2.3.1, only the minimal data fragments required for the current task are retrieved, with all processing performed on masked data while prohibiting access to any actual fields or values.

Within this framework, locally deployed system platforms are no longer limited to model inference but are evolving into internal hubs for model training and continuous optimization. By integrating multiple lightweight specialized models, enterprises can rapidly perform task-specific fine-tuning for subtasks such as intent recognition, SQL syntax correction, or schema alignment. Unlike cloud-based training, local platforms can access masked copies of enterprise data, enabling models to learn domain-specific logical and structural patterns within secure boundaries.

With the help of localized platforms, the AI Coding paradigm significantly lowers the technical barrier to data analysis by enabling domain experts to drive end-to-end analytical processes through natural language interaction, shifting the focus from traditional programming to intent-driven exe-cution. This framework leverages multi-agent orchestration to automate tasks such as SQL generation, visualization, and report synthesis, forming a complete analytical pipeline. However, this heavy reliance on AI-generated code introduces significant risks, including potential code quality flaws, security vulnerabilities, and data leakage, especially when handling sensitive information without robust oversight. It is crucial to strengthen validation and monitoring mechanisms to ensure the security and reliability of AI-assisted programming.

Current practice increasingly favors multi-model collaborative architectures: lightweight specialized models handle subtasks such as schema linking, syntax validation, or execution explanation, while LLMs focus on high-level reasoning and natural language interaction. These components communicate via standardized interfaces such as JSON Schema, forming a loosely coupled yet robust pipeline. Consequently, the Dual-Layer Model Architecture effectively supports the operational workflow of such multi-model structures. For instance, the CarbonChat system integrates "hallucination marking" and "generation verification" modules, cross-checking every analytical statement against its original evidence when generating SQL for carbon emission reports. Any unsupported inference is flagged and auto-corrected, ensuring that the final outputs are accurate, evidence-grounded, and decision-ready [50]. Similarly, closed-domain hallucination detection frameworks such as VeriTrail which provide traceability across arbitrary generation steps, are increasingly valued in enterprise intelligent agent applications [51].

## 5.2 Integration of Advanced Analytical Models

Although SQL provides well-established capabilities for structured data querying and execution, its limitations become apparent when applied to complex analytical tasks such as predictive modeling, anomaly detection, or semantic clustering. Traditional SQL is primarily designed for factual data retrieval and statistical aggregation and is not inherently suitable for expressing computational logic involving time-series modeling, nonlinear pattern recognition, or high-dimensional feature analysis. To address this gap, the current enterprise data systems are increasingly integrating SQL generation with AI or machine learning (ML) models, thereby extending analytical functionality from descriptive querying toward predictive and inferential analysis.

In this integrated framework, a natural language request is typically automatically decomposed by the system into multi-level tasks:

SQL Subtask: The SQL generation module translates the structured query requirements in the request into standard SQL statements for extracting the dataset, such as "sales data for the past six months".

ML Subtask: For the extracted data, call pretrained AI models for prediction, anomaly detection, cluster analysis, etc.

The data obtained by the SQL subtask specifically serves the AI module. The final results are presented in a unified frontend via visual charts, summary reports, or natural language explanations. Users do not need to distinguish the boundary between "query" and "prediction", the entire analysis process is automated end-to-end.

Taking the product KG as an example, which seamlessly combining structured KGs with generative AI models [48]. As shown in Figure 10, when users are interested in learning more about "Huawei TaiShan Server", they can right-click the node and select "AI Product Summary". The system immediately extracts the key information of the product such as name, brand and model, then calls the integrated third-party LLM in the background, and presents the detailed analysis report generated by the AI to the user in a pop-up window within seconds. The report content is comprehensive and professional, covering multiple dimensions such as product information overview, technical specifications, and application scenarios, greatly expanding the depth of the user's data analysis. This function successfully upgrades the KG from a "relationship displayer" to a "trigger for deep insights".

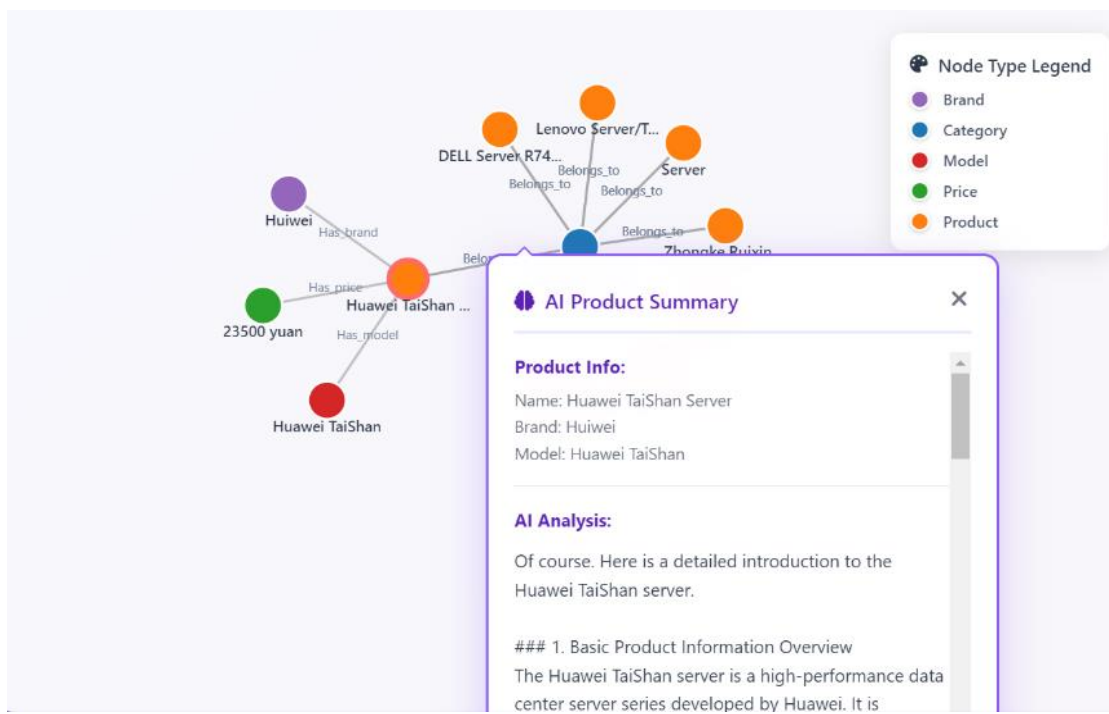

Figure 10. Example of integrating AI models into knowledge graph (cited and modified from [48]).

Another example, in the field of urban development research, the analysis of urban big data often requires screening and integration from multi-source heterogeneous datasets. Existing research has shown that machine learning and predictive models based on multi-dimensional urban data have achieved significant results in trend prediction and decision support [52][53]. However, the high heterogeneity of urban data, including differences in data definitions, inconsistencies in collection frequency and quality, and non-uniform spatial unit divisions, often makes it difficult for predictive models and indicator systems built for a single city or specific development region to fully match the actual situation. Furthermore, deviations in indicator systems and semantic definitions between different cities further increase the difficulty of model migration and generalization.

To address the aforementioned challenges, the LLM-based urban data intelligence framework is developed. This framework employs a SQL query agent that leverages SQL generation to perform semantic filtering and dynamic extraction from heterogeneous data sources. During the SQL subtask phase, a data profiling mechanism is integrated to automatically compute metrics such as missing rates, distribution characteristics, and temporal coverage, enabling adaptive selection of aggregation granularity and feature engineering strategies based on data properties. Subsequently, a modeling agent is called to conduct model construction and evaluation, supporting differentiated modeling and trend simulation tailored to the characteristics of individual urban datasets. This approach helps maintain semantic consistency while more effectively capturing dynamic urban patterns and regional heterogeneity, thereby providing technically grounded support for localized urban forecasting and policy analysis.

This hybrid querying paradigm not only expands the scope of SQL generation but also establishes a unified interface layer that bridges natural language, structured data, and intelligent analytics. Future research directions include: (1) developing multimodal parsers capable of interpreting analytical intent; (2) designing task decomposition and dynamic orchestration mechanisms for adaptive coordination between querying and reasoning; (3) exploring methods for model fusion and enhanced result interpretability under data security constraints. Along this integrative trajectory, SQL generation systems are evolving from linguistic interfaces to databases toward central coordinators of intelligent analytical workflows, positioning natural language as a viable interactive method for initiating complex knowledge generations.

## 6. Conclusion

This paper systematically studies the recent advances and practical system deployments of enterprise data analysis applications driven by LLMs and agent-based technologies. The main content includes schema linking, SQL generation models, multi-agent collaboration, KG augmentation, and real-world application scenarios. First, schema linking has evolved from simple keyword matching to an integrated approach that combines structured schema description generation, vector-based retrieval optimization, and domain knowledge injection. Methods such as M-Schema, KaSLA, and LitE-SQL enhance model comprehension of complex database schemas through refined context construction and example learning. Second, SQL generation models are shifting from reliance on closed-source LLMs toward open-source models using SFT and reinforcement learning techniques. SFT incorporates structured prior knowledge, while reinforcement strategies such as Direct Preference Optimization (DPO) and self-reward search, leverage SQL execution feedback for iterative refinement. These approaches enable open-source models to achieve performance on benchmarks like BIRD and Spider that is comparable to, or in some cases exceeds, the zero-shot performance of closed-source counterparts. Third, multi-agent frameworks address ambiguous queries and complex reasoning through role specialization and task decomposition strategies. Empirical studies suggest that the design of the workflow has a greater impact on system performance than the number of agents, concise and feedback-driven loops often outperform architectures with redundant intermediate steps. Fourth, KGs serve as external semantic layer that support query completeness verification, multi-turn interaction guidance, and improved few-shot example selection. They also help mitigate entity overfitting and provide structured priors to facilitate cross-domain transfer. In practical system deployments, enterprise applications must balance security, computational efficiency, and user experience. Multi-model tiered architectures help meet data compliance requirements. Collaborative paradigms like AI Coding lower the barrier for non-technical users. Besides, the integration of predictive models and vector retrieval extends the expressive capacity of SQL, enabling end-to-end pipelines that span querying, analysis, and decision support.

## References

[1]   Zhao, W. X., Zhou, K., Li, J., et al. (2023). A survey of large language models. arXiv preprint arXiv:2303.18223, 1(2).
[2]   Ge, Y., Mei, L., Duan, Z., et al. (2025). A Survey of Vibe Coding with Large Language Models. arXiv preprint arXiv:2510.12399.
[3]   Wenz, F., Bouattour, O., Yang, D., et al. (2025). BenchPress: A Human-in-the-Loop Annotation

System for Rapid Text-to-SQL Benchmark Curation. arXiv preprint arXiv:2510.13853.
[4] Chandra R, Agarwal S, Kumar S S, et al. OCEP: An Ontology-Based Complex Event Processing Framework for Healthcare Decision Support in Big Data Analytics[J]. arXiv preprint arXiv:2503.21453, 2025.
[5] Shi, L., Tang, Z., Zhang, N., et al. (2025). A survey on employing large language models for text-to-sql tasks. ACM Computing Surveys, 58(2), 1–37.
[6] Chakraborty, S., Pourreza, M., Sun, R., et al. (2025). Review, Refine, Repeat: Understanding Iterative Decoding of AI Agents with Dynamic Evaluation and Selection. arXiv preprint arXiv:2504.01931.
[7] Lei, F., Chen, J., Ye, Y., et al. (2024). Spider 2.0: Evaluating language models on real-world enterprise text-to-sql workflows. arXiv preprint arXiv:2411.07763.
[8] Li, J., Hui, B., Qu, G., et al. (2023). Can LLM already serve as a database interface? A big bench for large-scale database grounded text-to-sqls. Advances in Neural Information Processing Systems, 36, 42330–42357.
[9] Rall, D., Bauer, B., Mittal, M., et al. (2025). Exploiting Web Search Tools of AI Agents for Data Exfiltration. arXiv preprint arXiv:2510.09093.
[10] Wang, B., Shin, R., Liu, X., et al. (2019). RAT-SQL: Relation-aware schema encoding and linking for text-to-sql parsers. arXiv preprint arXiv:1911.04942.
[11] Lewis, P., Perez, E., Piktus, A., et al. (2020). Retrieval-augmented generation for knowledge-intensive NLP tasks. Advances in Neural Information Processing Systems, 33, 9459–9474.
[12] Liu, Y., Zhu, Y., Gao, Y., et al. (2025). Xiyan-SQL: A novel multi-generator framework for text-to-sql. arXiv preprint arXiv:2507.04701.
[13] Gladkykh, T., & Kirykov, K. (2025). Datrics Text2SQL: A Framework for Natural Language to SQL Query Generation. arXiv preprint arXiv:2506.12234.
[14] Lee, S. Y. T., Chen, J., Calzaretto, A., et al. (2025). VizCopilot: Fostering Appropriate Reliance on Enterprise Chatbots with Context Visualization. arXiv preprint arXiv:2510.11954.
[15] Gurawa, P., & Dharmik, A. (2025). Balancing Content Size in RAG-Text2SQL System. arXiv preprint arXiv:2502.15723.
[16] Yu, X., Jian, P., & Chen, C. (2025). TableRAG: A Retrieval Augmented Generation Framework for Heterogeneous Document Reasoning. arXiv preprint arXiv:2506.10380.
[17] Sergeev, A., Goloviznina, V., Melnichenko, M., et al. (2025). Talking to Data: Designing Smart Assistants for Humanities Databases. arXiv preprint arXiv:2506.00986.
[18] Dorbani, A., Yasser, S., Lin, J., et al. (2025). Beyond Quacking: Deep Integration of Language Models and RAG into DuckDB. arXiv preprint arXiv:2504.01157.
[19] Seabra, A., Cavalcante, C., Nepomuceno, J., et al. (2024). Dynamic multi-agent orchestration and retrieval for multi-source question-answer systems using large language models. arXiv preprint arXiv:2412.17964.
[20] Piao, S., Lee, J., & Park, S. (2025). LitE-SQL: A Lightweight and Efficient Text-to-SQL Framework with Vector-based Schema Linking and Execution-Guided Self-Correction. arXiv preprint arXiv:2510.09014.
[21] Yuan, Z., Chen, H., Hong, Z., et al. (2025). Knapsack optimization-based schema linking for LLM-based Text-to-SQL generation. arXiv preprint arXiv:2502.12911.
[22] Ma, X., Tian, X., Wu, L., et al. (2024). Enhancing text-to-sql capabilities of large language models via domain database knowledge injection. arXiv preprint arXiv:2409.15907.

[23] Peng, D. (2025). X-SQL: Expert Schema Linking and Understanding of Text-to-SQL with Multi-LLMs. arXiv preprint arXiv:2509.05899.
[24] Tritto, M., Farano, G., Di Palma, D., et al. (2025). GradeSQL: Outcome Reward Models for Ranking SQL Queries from Large Language Models. arXiv preprint arXiv:2509.01308.
[25] Hoang, C. D. V., Tangari, G., Lanfranchi, C., et al. (2025). Distill-C: Enhanced NL2SQL via Distilled Customization with LLMs. arXiv preprint arXiv:2504.00048.
[26] Parthasarathi, S. H. K., Zeng, L., & Hakkani-Tür, D. (2023). Conversational Text-to-SQL: An Odyssey into State-of-the-Art and Challenges Ahead. In ICASSP 2023-2023 IEEE International Conference on Acoustics, Speech and Signal Processing (ICASSP) (pp. 1–5). IEEE.
[27] Qin, Y., Chen, C., Fu, Z., et al. (2024). ROUTE: Robust multitask tuning and collaboration for Text-to-SQL. arXiv preprint arXiv:2412.10138.
[28] Yang, J., Hui, B., Yang, M., et al. (2024). Synthesizing text-to-SQL data from weak and strong LLMs. arXiv preprint arXiv:2408.03256.
[29] Zhong, V., Xiong, C., & Socher, R. (2017). Seq2SQL: Generating structured queries from natural language using reinforcement learning. arXiv preprint arXiv:1709.00103.
[30] Gajjar, K., Sikchi, H., Gautam, A. S., et al. (2025). CogniSQL-R1-Zero: Lightweight reinforced reasoning for efficient SQL generation. arXiv preprint arXiv:2507.06013.
[31] Yao, Z., Sun, G., Borchmann, L., et al. (2025). Arctic-Text2SQL-R1: Simple Rewards, Strong Reasoning in Text-to-SQL. arXiv preprint arXiv:2505.20315.
[32] Papicchio, S., Rossi, S., Cagliero, L., et al. (2025). Think2SQL: Reinforce LLM reasoning capabilities for text2sql. arXiv preprint arXiv:2504.15077.
[33] Lyu, S., Luo, H., Li, R., et al. (2025). SQL-o1: A Self-Reward Heuristic Dynamic Search Method for Text-to-SQL. arXiv preprint arXiv:2502.11741.
[34] Weng, H., Wu, P., Longjie, C., et al. (2025). Graph-Reward-SQL: Execution-Free Reinforcement Learning for Text-to-SQL via Graph Matching and Stepwise Reward. arXiv preprint arXiv:2505.12380.
[35] Ali, A., Baheti, A., Chang, J., et al. (2025). A State-of-the-Art SQL Reasoning Model using RLVR. arXiv preprint arXiv:2509.21459.
[36] Hao, H., Hu, W., Verkholyak, O., et al. (2025). PaVeRL-SQL: Text-to-SQL via Partial-Match Rewards and Verbal Reinforcement Learning. arXiv preprint arXiv:2509.07159.
[37] Kulkarni, A., & Srikumar, V. (2025). Reinforcing Code Generation: Improving Text-to-SQL with Execution-Based Learning. arXiv preprint arXiv:2506.06093.
[38] Lei, F., Meng, J., Huang, Y., et al. (2025). Reasoning-table: Exploring reinforcement learning for table reasoning. arXiv preprint arXiv:2506.01710.
[39] Gao, D., Wang, H., Li, Y., et al. (2023). Text-to-sql empowered by large language models: A benchmark evaluation. arXiv preprint arXiv:2308.15363.
[40] Li Z, Wang X, Zhao J, et al. Pet-sql: A prompt-enhanced two-round refinement of text-to-sql with cross-consistency[J]. arXiv preprint arXiv:2403.09732, 2024.
[41] Jehle, D., Purucker, L., & Hutter, F. (2025). Agentic NL2SQL to Reduce Computational Costs. arXiv preprint arXiv:2510.14808.
[42] Guo, T., Wang, H., Liu, C. C., et al. (2025). MTSQL-R1: Towards Long-Horizon Multi-Turn Text-to-SQL via Agentic Training. arXiv preprint arXiv:2510.12831.
[43] Heidari, O. R., Reid, S., & Yaakoubi, Y. (2025). AGENTIQL: An Agent-Inspired Multi-Expert Framework for Text-to-SQL Generation. arXiv preprint arXiv:2510.10661.

[44] Li, Y., Tao, R., Hommel, D., et al. (2025). Agent Bain vs. Agent McKinsey: A New Text-to-SQL Benchmark for the Business Domain. arXiv preprint arXiv:2510.07309.

[45] Lee, K., Hong, S., Park, J., et al. (2025). EMR-AGENT: Automating Cohort and Feature Extraction from EMR Databases. arXiv preprint arXiv:2510.00549.

[46] Guo, J., Patel, K., Ono, J. P., et al. (2025). Rethinking Agentic Workflows: Evaluating Inference-Based Test-Time Scaling Strategies in Text2SQL Tasks. arXiv preprint arXiv:2510.10885.

[47] Xue, S., Jiang, C., Shi, W., et al. (2023). DB-GPT: Empowering database interactions with private large language models. arXiv preprint arXiv:2312.17449.

[48] Wang X, Ling X, Li K, et al. Multi-dimensional Data Analysis and Applications Basing on LLM Agents and Knowledge Graph Interactions[J]. arXiv preprint arXiv:2510.15258, 2025.

[49] Han, H., Wang, Y., Shomer, H., et al. (2024). Retrieval-augmented generation with graphs (GraphRAG). arXiv preprint arXiv:2501.00309.

[50] Cao, Z., Han, M., Wang, J., et al. (2025). CarbonChat: Large Language Model-Based Corporate Carbon Emission Analysis and Climate Knowledge Q&A System. arXiv preprint arXiv:2501.02031.

[51] Metropolitansky, D., & Larson, J. (2025). VeriTrail: Closed-Domain Hallucination Detection with Traceability. arXiv preprint arXiv:2505.21786.

[52] Wang, X., Li, X., Wu, T., He, S., Zhang, Y., Ling, X., Chen, B., Bian, L., Shi, X., Zhang, R., et al. (2024). Municipal and Urban Renewal Development Index System: A Data-Driven Digital Analysis Framework. Remote Sensing, 16(3), 456. https://doi.org/10.3390/rs16030456

[53] Wang, X., Chen, B., Li, X., Zhang, Y., Ling, X., Wang, J., Li, W., Wen, W., & Gong, P. (2022). Grid-Based Essential Urban Land Use Classification: A Data and Model Driven Mapping Framework in Xiamen City. Remote Sensing, 14(23), 6143. https://doi.org/10.3390/rs14236143